# A Data Augmentation Method for Fully Automatic Brain Tumor Segmentation

WANG Yu*, JI Yarong, XIAO Hongbing

(Beijing Key Laboratory of Big Data Technology for Food Safety, School of Artificial Intelligence,

Beijing Technology and Business University, Beijing 100048, China)

**Abstract**: Automatic segmentation of glioma and its subregions is of great significance for diagnosis, treatment and monitoring of disease. In this paper, an augmentation method, called TensorMixup, was proposed and applied to the three dimensional U-Net architecture for brain tumor segmentation. The main ideas included that first, two image patches with size of 128×128×128 voxels were selected according to glioma information of ground truth labels from the magnetic resonance imaging data of any two patients with the same modality. Next, a tensor in which all elements were independently sampled from Beta distribution was used to mix the image patches. Then the tensor was mapped to a matrix which was used to mix the one-hot encoded labels of the above image patches. Therefore, a new image and its one-hot encoded label were synthesized. Finally, the new data was used to train the model which could be used to segment glioma. The experimental results show that the mean accuracy of Dice scores are 91.32%, 85.67%, and 82.20% respectively on the whole tumor, tumor core, and enhancing tumor segmentation, which proves that the proposed TensorMixup is feasible and effective for brain tumor segmentation.

## 1 Introduction

Glioma is one of the most commonly primary brain tumors which grows from glial cells and is highly aggressive. Generally, two types of the tumor including high grade gliomas (HGG) and low grade gliomas (LGG) exist. Patients with LGG can usually survive for 5 years or longer, while patients with HGG can only survive for 2 years or even less [1]. Locating the tumor accurately is difficult even for experts, not only because of its variabilities in position, shape, and size, but also due to its vague boundary with surrounding normal tissues. In recent years, medical imaging technologies such as computed tomography scans or magnetic resonance imaging (MRI) scans have been used extensively in clinical treatment. Among them, MRI as a typically non-invasive imaging technology, can produce high quality brain images without injury and skull artifacts, and can provide more comprehensive information in the form of multi-modalities. In addition, MRI has been the main technical method of diagnosis and treatment of brain tumor [2].

Traditional diagnosis of brain tumors mainly relies on doctors by reading MRI images, which leads to heavy burden for doctors. Therefore, using brain tumor segmentation algorithms to automatically detect brain tumors of patients by computers has been the focus of research in recent years. In this study, the segmentation algorithm takes MRI images as input and classifies the voxels to four sub-categories including edema, necrosis & non-enhancing tumor, enhancing tumor, and background. Consequently, the algorithm is evaluated mainly based on its segmentation accuracy of the whole tumor (WT), tumor core (ET) and enhancing tumor (ET). To be specific, the WT is composed of edema, necrosis, non-enhancing tumor and enhancing tumor tissues. While the TC includes necrosis, non-enhancing tumor, and enhancing tumor tissues, and the ET denotes enhancing tumor tissues.

At present, a large number of deep learning algorithms have been proposed with the great development of computer technologies, especially the convolutional neural network (CNN) based on supervised learning has been widely used in the segmentation of brain tumor, and can achieve state-of-the-art segmentation results. Havaei et al. [3] proposed a two-pathway brain tumor segmentation network to extract multi-scale features, and to predict category of central voxel in each image patch. Shen et al. [4] constructed a symmetry-driven network based on fully convolutional network (FCN) to achieve densely classification for brain images.

In 2015, Ronneberger et al. [5] proposed U-Net based on FCN for biomedical image segmentation. This network is widely applied due to the deep fusion of high and low level features brought by the structure of U-shaped encoder-decoder and skip connection. Its encoder fully extracts high-level semantic features from the input image, while the decoder uses up-sampling operations to convert low-resolution features extracted from the encoder into high-resolution features, which can achieve end-to-end segmentation and can identify the location and structure of the region of interest (ROI). In addition, the U-Net also uses skip connection structure to combine high and low level features respectively resulted from the decoder and encoder at the same level, which makes up the defects about how much detailed information is lost in images caused by the up-sampling operations, and effectively improves the segmentation accuracy of the model. Dong et al. [6] applied two dimensional (2D) U-Net to brain tumor segmentation tasks. However, 2D network cannot fully extract the contextual information of MRI data between slices. To handle this problem, Beers et al. [7] proposed a brain tumor segmentation model with three dimensional U-Net (3DUNet) which can effectively improve the segmentation accuracy.

Class imbalance problem usually appears in the research of brain tumor segmentation based on deep learning, because the volume of the WT in MRI images of patients is much smaller than that of the whole brain, and generally accounts for only 1.54% of the entire image [8], while the volume of the TC and ET occupies even less, which will make the model expose to more negative rather than positive sample features during the training process. Therefore, it is detrimental to fully learn the tumor features, especially for TC and ET, and is further bad for the improvement of accuracy for the segmentation model. To solve this problem, Wang et al. [9] proposed an anisotropic CNN model which used three cascaded FCN networks to segment the WT, TC and ET region, and degraded multiple classification tasks into a series of binary classification tasks, and simplified the overall architecture. Li et al. [10] and Mckinley et al. [11] adopted the Focus Loss function to reduce the weight of negative samples, so that the model could focus more on the features of positive samples during training, thus alleviating the class imbalance problem.

Moreover, the construction of brain tumor segmentation model needs a large number of annotated image data to train the deep neural network. The annotated data, however, is scarce at present. The reason lies in that the annotation of medical images generally needs to be completed by experienced professionals who are scarce. At the same time, annotation of medical images is a difficult, time-consuming and expensive work. Thus annotated data is often insufficient for training of models. Aiming at this problem, data augmentation techniques are used to produce more diverse data by applying certain changes to the original data, which can effectively prevent the model from over-memorizing the data. Basic data augmentation methods include flipping, rotating, panning, scaling, changing brightness, adding Gaussian noise and elastic distortion etc. However, in some study it was found that when basic data augmentation methods are applied to the training of the tumor segmentation model, only a minor improvement appears, and the conclusion of the literature [12] also confirmed the above find.

In recent years, Mixup, a new technique based on the data interpolation [13], was proposed for the data augmentation of image classification tasks. This method mixes two randomly selected images in the way of convex combination, and combines its one-hot encoded labels in the same way. Thus a new image and its label can be synthesized. Using newly synthesized data to train deep learning model can alleviate the issue of deficiency of labeled data. Subsequently, Mixup can be applied to the research of medical image segmentation. Based on this method, Zach et al. [14] proposed mixmatch method to augment data in which the samples were matched with specific criteria before mixed, then were synthesized into new samples. The related results demonstrated that it could effectively improve model performance by applying Mixup to brain tumor segmentation. Yin et al. [15] proposed the Tumor Mixup method which firstly used masks to process the brain images of two patients. Then the Mixup method was used to mix the two masked images. So the synthesized image data was generated and contained tumor features of both patients, which made it possible that the model could learn the tumor features of two patients in one image and could improve the training efficiency of the model. In addition, Guo [16] further applied Mixup to mixing text data for text classification task by using a matrix in which all elements were independently sampled from *Beta*($\alpha$, $\alpha$) distribution, which could improve the classification accuracy of the model in a nonlinear way.

Inspired by the above ideas, in this paper a new data augmentation technique, called TensorMixup, is proposed, in which two image patches with fixed size containing tumor regions are obtained respectively from the MRI images of two randomly selected patients. Then a tensor is used to mix these two image patches. Next a matrix mapped from this tensor is used to mix two one-hot labels of the above image patches. So the new labeled data is synthesized and applied to training model. The main contributions of this work are as follows: 1) The porposed TensorMixup method mixes the image patches containing much tumor features rather than whole images, so that each mixing process occurs in the ROIs of images. 2) TensorMixup uses a tensor to mix image patches, so that each pair of voxels in the pair of image patches is assigned an independent mixing strategy, instead that all pairs of voxels share the same mixing strategy like traditional Mixup method. So the proposed algorithm could produce more diverse training data by mixing in a nonlinear way and could make it more continuous for sample points in the input space. 3) The tensor utilized to mix is mapped to a matrix which can mix the one-hot encoded sequences. Thus the problem that the tensor cannot directly mix the one-hot labels of image patches can be solved. The results of experiment show that the proposed algorithm can effectively alleviate the problems of insufficient data and class imbalance of data in the brain tumor segmentation task. Training efficiency can also be improved simultaneously.

## 2 Method

### 2.1 Data Introduction

The dataset is provided by Multimodal Brain Tumor Image Segmentation Challenge

(BraTS) [17] of Medical Image Computing and Computer Assisted Intervention Society (MICCAI). More specifically, the BraTS 2019 dataset contains 335 patients, including 259 patients with HGG and 76 patients with LGG, and the BraTS 2015 dataset contains 186 patients, including 132 patients with HGG and 54 patients with LGG. All data have been pre-processed with skull-stripping, image registration and spatial normalization by the challenge organizers. The data for each patient includes four modal MRI images which are T1-weighted MRI (T1), T1-weighted MRI with contrast enhancement (T1ce), T2-weighted MRI (T2) and Fluid-Attenuated Inversion Recovery (Flair) and one ground truth label. Different modal images can provide different information about the tumor. The ground truth label annotated by experts potentially contains four categories, namely background, necrosis & non-enhancing tumor, edema and enhancing tumor. All images have the same size with 240×240×155 voxels. The data of a patient is shown in Fig.1.

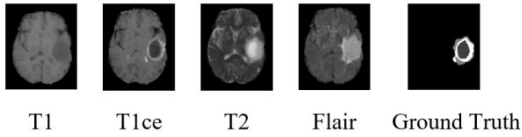

Fig.1 Four modal MRI images and the ground truth of a patient

## 2.2 Data Preprocessing

The data provided by BraTS has only been pre-processed preliminarily. For building the state-of-art model, the bias field correction and standardization are also done in this study. MRI images have a property on intensity inhomogeneity which is also called bias field effect and is caused by the inhomogeneity of magnetic field or the slight movements of patients when patients are scanned by equipment for acquiring MRI images. So the N4ITK bias field correction method is used in this study to remove this inhomogeneity in MRI images [10]. In addition, because the data is collected from different equipment and institutions, the ranges of values in different data are inconsistent, so the Z-Score standardization method is used to unify the values of all data into a same and smaller range. Thus the values of the images present a normal distribution, which is conducive to numerical calculation of the model in training process. Standardization is shown in Equation 1 where $X$ denotes the intensity matrix of one modal image, $\overline{X}$ is the average value of $X$, $X_{std}$ means the standard deviation of $X$.

$$X = \frac{X - \overline{X}}{X_{std}} \qquad (1)$$

## 2.3 TensorMixup Data Augmenta-tion Method

In this paper, aiming at expanding the types and quantities of experimental data, a data augmentation method, called TensorMixup, is proposed and used for the brain tumor segmentation task.

### 2.3.1 Traditional Mixup Data Augmentation Method

Mixup was proposed by Zhang et al. [13] based on the principle of vicinal risk minimization [18] and was initially applied to image classification tasks. This method performs linear convex combination of the input $x_i$ and $x_j$ from the sample pair $(x_i, y_i)$ and $(x_j, y_j)$, as well as their corresponding labels $y_i$ and $y_j$, which can synthesize a new sample $(x, y)$ [19]. The synthetic samples generated in this way are located in the vicinal of original samples and are fed into the neural network for training model. The process of using Mixup to generate new samples is described by:

$$x = \lambda * x_i + (1-\lambda) * x_j \qquad (2)$$

$$y = \lambda * y_i + (1-\lambda) * y_j, \qquad (3)$$

where $x_i$ and $x_j$ denote the inputs of two original samples, while $x$ denotes the input of the synthesized sample. For Equation (3), $y_i$ and $y_j$ denote the one-hot encoded labels of $x_i$ and $x_j$ respectively, while $y$ denotes the one-hot encoded label of $x$. In addition, $\lambda$ is a scalar randomly sampled from *Beta* ($\alpha$, $\alpha$) distribution in the range of [0, 1] and $\alpha$ is a hyper-parameter in the range of [0, +∞) which controls the distribution of Beta. Zhang et al. [13] confirmed that when $\alpha$=0, Mixup is not used. When $\alpha$=1, *Beta* ($\alpha$, $\alpha$) distribution is equivalent to 0-1 distribution. When $\alpha$>1, *Beta* ($\alpha$, $\alpha$) distribution is similar to the normal distribution. And $\alpha$→+∞ means that the value of $\lambda$ is constant to 0.5. So using *Beta* $\alpha$, $\alpha$) distribution to generate mixing ratio is pretty flexible. For Mixup, the two inputs as well as the two labels in two samples are added with the weights of $\lambda$ and 1-$\lambda$, so that the new generated sample is between the two original samples. Compared to apply only true data for model training, using the new sample and the original sample together to fit the objective function can make the function locally tend to linearization [20] and can prevent the function from overfitting.

To alleviate the problem of insufficient data in this work, Mixup was initially used for the brain tumor segmentation. Specifically, given $X_i$ and $X_j$, the MRI brain images of two patients in the same modality, and their corresponding one-hot encoded labels are $Y_i$ and $Y_j$. Based on the Equation (4)-(5), the new image $X_s$ and its one-hot encoded label $Y_s$ are generated, and the scaler $\lambda$ is still sampled from the $Beta(\alpha, \alpha)$ distribution. At the end of the mixing process, the image patch with size of 128×128×128 voxels is selected from the synthetic data $(X_s, Y_s)$ and is sent into the neural network for model training.

$$X_s = \lambda * X_i + (1-\lambda) * X_j \quad (4)$$

$$Y_s = \lambda * Y_i + (1-\lambda) * Y_j \quad (5)$$

### 2.3.2 The Proposed TensorMixup Data Augmentation Method

Similar to Mixup, the proposed TensorMixup can also mix the information of two samples to generate new training data. However, unlike Mixup which mixes the whole images directly, TensorMixup mixes two image patches called tumor image patches whose tumor feature quantities are much more than feature ones of normal tissues. So the mixing process only occurs in the ROIs of original images，which can improve the mixing efficiency and can alleviate the class imbalance problem. Furthermore, TensorMixup assigns a different scaler as the mixing policy for each pair of voxels during mixing two MRI images, while Mixup uses a scaler for the mixing of every pair of voxels. Thus the proposed algorithm can generate more diversely synthetic data in the non-linear mixing way. The three steps need to be completed when TensorMixup is applied to the brain tumor segmentation. They are in order the selection of tumor image patches, the mixing of tumor image patches and the mixing of one-hot encoded labels. The three steps will be described in detail below.

**(1) Selection of Tumor Image Patches**

TensorMixup mixes the ROIs of two images. As mentioned above, Mixup has been applied to brain tumor segmentation tasks. However, a serious class imbalance problem exists when two whole images are directly mixed to generate new image data. The reason is that the volume of tumor region is far smaller than that of the whole brain in raw data, and it is not consistent for the locations of tumor regions in two images. Therefore, tumor features will be mixed with background features in all likelihood, resulting that the region containing tumor features may also contain a large amount of background features in synthetic images. In new data the number of tumor features is further reduced, and the class imbalance problem is aggravated, which is bad for the implement of the model. To avoid the above situation, only tumor regions in a pair of image data are mixed in TensorMixup method. At the same time, it is necessary to select the tumor image patches containing ROI from two original images.

The detailed procedure for obtaining tumor image patches is as follows. Firstly, two MRI images (240×240×155 voxels) with the same modality from two randomly selected patients are pre-processed. The pre-processing operations are described in section 2.2. Then the image patches containing ROI are obtained from pre-processed images. The execution of this process mainly depends on the boundary box information of tumor regions which could be acquired from the ground truth labels among original images. Next, each dimension of the image patches need be resized so as to be greater than or to be equal to 128. Because a certain dimension of the boundary boxes for tumor regions may be less than 128 and the size of input images of the deep neural network used in this paper is 128×128×128 voxels, the image patches which do not meet the requirement need be filled with zeros so that each dimension of image patches is 128 at least. Finally, the tumor image patches with size of 128×128×128 voxels are obtained from the above processed image patches. Similarly, the labels are also processed as the above operations.

**(2) Mixing of Tumor Image Patches**

A tensor $\varLambda$ is used to mix the information of tumor image patches after they are acquired from the original image pair, and a new image data is synthesised in TensorMixup method. The mixing process of patches is shown in equation (6) where the tumor image patches are denoted as $X1$ and $X2$, and the new generated image patch is denoted as $X$. The size of tumor image patches in equation (6) are 128×128×128 voxels. $\varLambda$ is a tensor in which all elements are sampled from $Beta$ ($\alpha$, $\alpha$) distribution, and the tensor's size is the same as one of tumor image patches. The symbol '$\odot$' denotes the Hadamard product operation. The process of composing a new image patch using TensorMixup is shown in Fig.2 in which the original images are the Flair modalities of two patients. In order to display lesion regions more clearly, the image was presented in the form of hot map.

$$X = \varLambda \odot X1 + (1-\varLambda) \odot X2 \qquad (6)$$

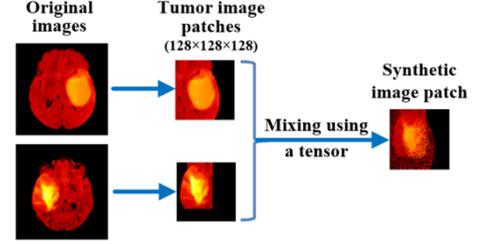

Fig. 2 The process of mixing using TensorMixup

Both in TensorMixup and Mixup, the mixing process at the voxel level is same in which a scalar is used to perform linear convex combination for two voxels at the same position in the image pair, and to generate new voxel values. In Mixup method only one scalar $\lambda$ is used to mix two whole images each time, which makes each voxel pair to be fused with the same mixing weight of $\lambda$. While in proposed TensorMixup method, different voxel pairs are fused with different mixing weights. Because mixing weights are the values of elements in tensor $\varLambda$ which are randomly sampled from $Beta(\alpha, \alpha)$ distribution. Thus more diverse images can be synthesized. Supposing only two images are used for training model, and are denoted as points A and B in the sample space as shown in Fig.3. When the A and B are mixed using Mixup, the point of the synthetic image can only locates on the line segment AB. When the above mixing process is repeated with different scalar, all synthetic data points appear on the line segment AB, and look gradually continuous. However, the other neighborhood spaces around segment AB are not filled with synthetic data points. But by TensorMixup, the synthetic data points generated could appear in the neighborhood spaces around segment AB besides on the line segment AB due to the nonlinear mixing method on TensorMixup in which different voxel pairs use different mixing weights.

So TensorMixup can make up the defects of Mixup in data expansion, and can make the input space of model further continuous.

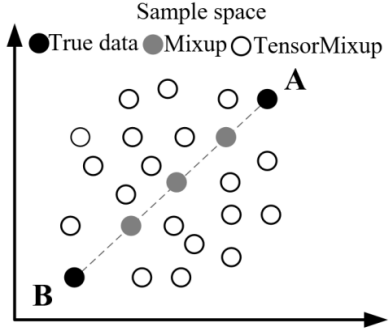

Fig. 3 Distribution map of data synthesized by Mixup and TensorMixup in sample space

(3) **Mixing of One-Hot Encoded Labels**

The segmentation network trained in this paper is a classification model where the labels of inputs need to be converted into one-hot encoded sequences to further calculate the loss function. During the above course the element values in ground truth labels must be integers. However, the voxels of synthetic labels often are not be integers if two labels of *X1* and *X2* are directly mixed according to equation (6), which would prevent the synthetic labels from being converted to one-hot encoded sequences, and would affect the calculation of the loss function. To avoid this problem, TensorMixup first transforms the labels of *X1* and *X2* into one-hot encoded sequences *Y1* and *Y2*. Then *Y1* and *Y2* are mixed to generate the one-hot encoded label *Y* of the new image *X*.

The specific process of transform for labels of *X1* and *X2* is as follows. Firstly, the labels of *X1* and *X2* are reshaped as a one-dimensional vector. Then each element in this vector is encoded as a *k*-dimensional binary vector which is the one-hot encoded label of a voxel corresponding to the element in *X1* or *X2*, and *k* represents the number of categories predicted by the model. Here, the value of *k* is 4 since the target of the research is a four classification task. As a result, the labels of *X1* and *X2* are encoded into two dimensional (2D) matrixes *Y1* and *Y2*, both of which have size of $128^3 \times 4$ and the number of rows in matric is the total numbers of voxels that need to be predicted. In addition, each row of *Y1* and *Y2* represents a one-hot encoded label of a voxel in *X1* or *X2*, and in each row the four elements represent the true probabilities that the voxel corresponding to the row in *X1* or *X2* belongs to four categories.

The one-hot encoded labels of tumor image patches are mixed after they are acquired. Because the size of *Y1* and *Y2* is inconsistent with that of tensor *Λ* and **1-*Λ***, tensor *Λ* cannot be used for the mixing of *Y1* and *Y2*. For solving this problem, the tensor *Λ* is mapped into *Λ\** according to the way by which ground truth labels are transformed into one-hot encoded sequences. Then the matrix *Λ\** is used for mixing label. The process of mapping from *Λ* into *Λ\** is as follows. Firstly, *Λ* is transformed into a one-dimensional column vector which is denoted as $Λ_v$ where the *i*th ($i \in [0, 128^3)$) element *m* ($m \in Λ_v$) is the mixing weight applied to the *i*th voxel of *X1* and the *i*th row vector of *Y1*. Then $Λ_v$ is used to construct *Λ\** with size of $128^3 \times 4$ as equation (7), in order to make the elements of *Y1* and *Y2* correspond to those of *Λ\** and (1-*Λ\**) during the mixing calculation. The relation of *Λ\** and $Λ_v$ is shown in Equation (8) where $vec(.)$ represents the vectorization operation of the tensor *Λ*.

$$Λ^* = f(Λ_v) = [Λ_v, Λ_v, Λ_v, Λ_v] \quad (7)$$
$$Λ_v = vec(Λ) \quad (8)$$

After the original labels and the tensor *Λ* are transformed respectively, the label data would be mixed, and the mixing process is shown in Equation (9). When the new data (*X*, *Y*) are synthesized, it can be directly feed into the neural network to minimize the loss, and can be used to optimize the parameters of the model.

$$Y = Λ^* \odot Y1 + (1 - Λ^*) \odot Y2 \quad (9)$$

## 2.4 Model Training

In this paper the synthetic data and real data are used to train the model jointly to achieve fully automatic segmentation model of brain tumors.

### 2.4.1 3DUNet

The 3DUNet used in this paper is composed of encoded pathway and decoded pathway. The encoded pathway is used to extract high-level semantic information of input images, and the decoded pathway relocates the structure of the region of interest with the help of extracted features. The encoded pathway consists of five context modules. Each module contains two $3\times3\times3$ convolutional layers with a stride of 1 and a dropout layer. The $3\times3\times3$ convolution layers with a stride of 2 and residual structures are used to connect context modules. The decoded pathway is mainly composed of position modules and upsampling modules. The position module contains two $3\times3\times3$ convolution layers with a stride of 1, and the upsampling module includes one upsampling layer and one $3\times3\times3$ convolution with a stride of 1. Four upsampling operations are carried out, so that low-resolution feature maps extracted from the encoded pathway are converted into high-resolution feature maps increasingly until they reach the same size as input images. So this architecture can achieve the goal of end-to-end segmentation. On the other hand, the deep supervision is utilized by adding three segmentation layers to the network and summing outputs of each segmentation layer to generate the final feature map which is feed into the softmax layer to obtain the probability map as the output of model. The segmentation layer is $1\times1\times1$ convolution with a stride of 1 which is mainly uesd to fuse information of different channels at the same position, and to reduce channel numbers of feature maps without changing the spatial resolution [21]. The 3DUNet architecture is shown in Fig. 4.

### 2.4.2 Joint Training

The training process mainly consists of two stages. In the first stage, the data synthesized by TensorMixup in each iteration is fed into the model in real time and the initial learning rate is 1e-03. The training of this stage will be finished when the segmentation accuracy of the model in the testing dataset is no longer improved. In the second stage, the real data which has been pre-processed is used to optimize the model obtained in the first stage until this model reaches the best performance. The initial learning rate is still 1e-04 and the basic data augmentation is carried out on the real data to prevent the model from over-fitting in later training period of this stage. During the training of two stages, the model updates network weights using Adam [22] optimizer with a decay rate of 1e-05, the activation function used is Leaky ReLU [23] and the normalization method is instance normalization, at the same time, the batch size is set to 2 and the dropout value is configured to 0.5. The 190 epochs will be done to achieve the state-of-art model and the Focus Loss is used to minimize the loss value.

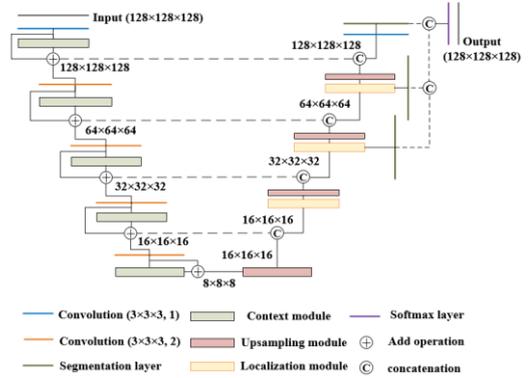

Fig. 4 The architecture of 3DUNet

## 3 Results and Analysis

### 3.1 Evaluation Methods

In this paper, the Dice similarity coefficient (Dice), Sensitivity, Specificity and Hausdorff distance are used to evaluate the performance of the model. Dice coefficient refers to the proportion of the intersection area to the total area between tumor regions predicted by the model and real tumor regions, which is the most comprehensive evaluation index and is taken as the main standard of the measurement in this paper. Sensitivity is the ratio of tumor voxels predicted correctly to the total tumor voxels, and Specificity refers to the ratio of background voxels predicted correctly to the total background voxels. Hausdorff distance is used to measure the surface distance between the tumor area predicted and the real lesion. In this paper 95% Hausdorff distance is used to measure model segmentation accuracy of the boundary. The expressions of four indexes are shown in Equation (10)-(13):

$$Dice(P,T) = \frac{|P \wedge T|}{(|P|+|T|)/2} \quad (10)$$

$$Sensitivity(P,T) = \frac{|P \wedge T|}{|T|} \quad (11)$$

$$Specificity(P,T) = \frac{|P_0 \wedge T_0|}{|T_0|} \quad (12)$$

$$Hausdorff(X,Y) = max\{\max_{x \in X}\min_{y \in Y} d(x,y), \max_{y \in Y}\min_{x \in X} d(x,y)\}, \quad (13)$$

where $P$ and $P_0$ respectively denotes the tumor and background area predicted by the model. $T$ and $T_0$ correspondingly are the real tumor and background area. '$\wedge$' represents a logical 'and' operation and '|.|' represents the size of sets. The $x$ is the point on the surface $X$ of the region $T$, and $y$ is the point on the surface $Y$ of the region $P$. The function $d(.)$ is used to calculate the distance between the points $x$ and $y$.

### 3.2 Experimental Settings

In this paper, the model is built on Ubuntu16.04 operating system, Pytorch and Python3.7, and Nvidia RTX 2080 Ti GPU is used to realize the algorithm calculation. The datasets used in this experiment are BraTs2019 and BraTs2015. For each dataset, five-fold cross-validation is used to obtain the optimal model and parameters. Specifically, first the dataset is divided into five groups, every time one group is selected as a testing dataset to test the model performance and remaining four groups are used as a training dataset to optimize the model. So a total of five kinds of training-testing datasets can be selected according to the principle of non-repetition in this way. Then the five kinds of datasets are used for five turns of training respectively and the model will be re-initialized for each turn. Finally, the model with the best test results is selected when five turns of training are completed [24].

To save computational resources and to reduce memory occupation, the size of input in the training stage is 128×128×128 voxels. In the first stage of training, the new image patch is synthesized by TensorMixup and is taken as the input of the model. In the second stage, the real data are preprocessed using basic data augmentation methods including flipping (horizontal and vertical), rotation (90°, 180° and 270°), adding Gaussian noise, changing brightness, and elastic distortion. Then the image patches with size of 128×128×128 voxels are taken out from preprocessed images and are fed into the model. The acquisition of image patches is mainly based on the ground truth data to obtain the information of tumor boundary boxes, and the margin of the boundary box is set to 3. In the testing stage, the input images with the size of 240×240×160 voxels are directly fed into the model. These inputs are generated by only filling 0 around testing images (240×240×155), so they can fit the convolution operation in the forward propagation. After the output is obtained, the segmentation result of testing images will be extracted from the output and be used to evaluate the performance of the model.

The value of parameter $\alpha$ determines the distribution of elements of the tensor $\Lambda$. In order to select an appropriate value about $\alpha$, BraTs2015 Dataset is divided into training set and testing set with a ratio of 4:1. Specifically, $\alpha$ was set to 0.3, 0.4, 0.5, 0.6, 1 and 1.2 respectively under the condition in which other variables were constant. So the six groups of models were trained successively, and 120 training epochs were carried out in each group. The mean Dice scores of six models in WT varying with iterations are as shown in Fig.5. The final evaluation results of six models in mean Dice index are shown in Tab. 1.

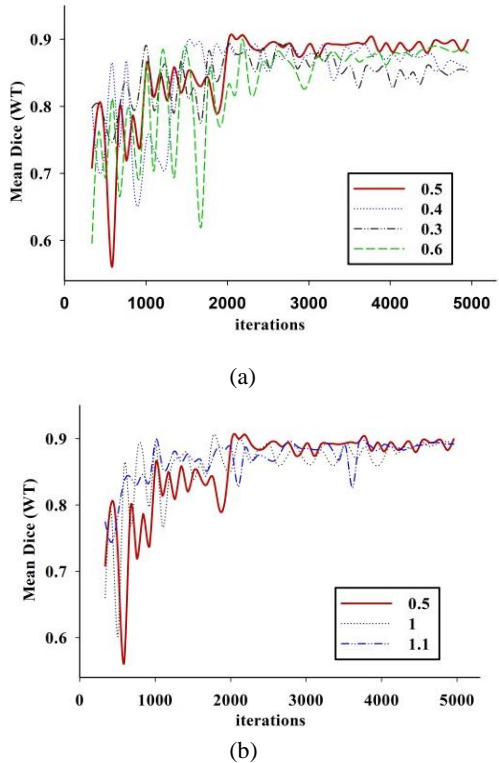

(a)

(b)

Fig. 5 Mean dice scores in WT on BraTs2015 testing dataset using proposed method with $\alpha$ of (a) 0.3, 0.4, 0.5 and 0.6, and (b) 0.5, 1 and 1.1.

It can be seen from Fig. 5(a) and 5(b) that the mean Dice scores of the model with $\alpha$ of 0.5 is obviously better than those of other $\alpha$ values. At the same time, when α is set to 0.5, the mean Dice scores of the model in WT are close to or reach 0.9 many times, and their fluctuation range is relatively small. In addition, there is still an upward trend at the late training for the model with α of 0.5. In Tab.1 the model with α of 0.5 has a good segmentation performance overall. Therefore, the parameter $\alpha$ is set to 0.5 for subsequent experiments.

Tab. 1 Results of six models with different $\alpha$ values in mean Dice index (%)

| $\alpha$ | WT | TC | ET |
|---|---|---|---|
| 0.3 | 89.22 | 85.27 | **81.28** |
| 0.4 | 89.22 | 84.29 | 79.58 |
| 0.5 | **89.74** | 84.99 | 81.16 |
| 0.6 | 88.99 | 81.52 | 79.53 |
| 1 | 88.54 | 84.22 | 78.92 |
| 1.1 | 89.28 | **85.49** | 78.05 |

## 3.3 Results and Analysis

### 3.3.1 Effectiveness of TensorMixup

To verify the effectiveness of the proposed algorithm, the 3DUNet was trained on BraTs2019 dataset using TensorMixup (TM), Mixup (M) and Base Augmentation (BA) methods respectively. Besides different augmentation methods were used, the other experimental settings were the same for the three models. All of them were trained with 190 epochs and the size of their inputs is 128×128×128 voxels. When training the 3DUNet with M model (3DUNet+M), Mixup was used to mix pre-processed images to generate a new image where an image patch with fixed size was taken out and was fed into the model. When training the 3DUNet with BA model (3DUNet+BA), the basic data augmentation methods were operated. Then the input image patches were acquired from transformed images. The testing results of three models were shown in Tab. 2.

Tab. 2 Experimental results of models using different data augmentation methods on BraTs2019 testing dataset

| Model | Mean Dice (%) | | | Mean Sensitivity (%) | | | Mean Hausdorff (mm) | | | Mean Specificity (%) | | |
|---|---|---|---|---|---|---|---|---|---|---|---|---|
| | WT | TC | ET | WT | TC | ET | WT | TC | ET | WT | TC | ET |
| 3DUNet+BA | 89.79 | 85.13 | 80.90 | **94.08** | **87.46** | 80.77 | 13.61 | **7.47** | 5.45 | 99.84 | 99.94 | **99.96** |
| 3DUNet+M | 90.55 | 85.46 | 81.61 | 93.54 | 85.41 | **88.22** | **10.06** | 9.98 | 6.48 | 99.89 | 99.96 | 99.94 |
| **3DUNet+TM** | **91.32** | **85.67** | **82.20** | 90.23 | 85.88 | 87.47 | 10.88 | 8.71 | **4.73** | **99.93** | **99.96** | 99.95 |

It can be seen from Tab. 2 that the 3DUNet with TM (3DUNet+TM) model gets mean Dice scores of 91.32%, 85.67% and 82.20% respectively in WT, TC and ET, which are the optimal performance compared with other models, and it demonstrates that the proposed TensorMixup is an effective data augmentation technique for brain tumor segmentation tasks. Among the three models, when 3DUNet+TM model is used to segment ET, its mean Dice score reaches the highest and mean Hausdorff distance reaches the lowest, and its Sensitivity was only lower than the optimal one by 0.75%. It shows that the proposed method is beneficial to the segmentation of ET region and its boundary.

In additional, the overall evaluation results of the 3DUNet+TM and 3DUNet+M model are better than that of model trained only using basic data augmentation methods. 3DUNet+TM and 3DUNet+M models have higher Dice values than the 3DUNet+BA model in all three regions by more than 0.33%. The above results verify that the mixing algorithms have practical significance for data augmentation in brain tumor segmentation tasks, and are more beneficial than basic data augmentation methods for improving the segmentation performance of model to some extent.

To analyze the overall segmentation capability of the proposed algorithm, in this study TensorMixup model is used to segment all testing data of BraTs2019 and the segmentation results are evaluated. The experimental results of all testing data were presented as shown in Fig. 6. It can be seen from Fig. 6 (a) that the Dice scores of all testing data in WT can basically exceed 0.9, and the Dice scores of TC and ET can be more than 0.8. The median of Dice scores in TC can reach 0.9 and its mean is only around 0.82. This indicates that the accuracy of TensorMixup model when predicting TC is generally around 0.9 and is also relative low when accessing to few abnormal data, thus resulting in a large gap between the mean and the median of Dice scores in TC. Fig. 6(b) shows that the Sensitivities when TensorMixup model is used to predict WT and TC are about 0.9 and 0.86 respectively, and the Sensitivities in ET is roughly between 0.83~0.96. Fig. 6(c) reveals that the median of Hausdorff distance in WT is less than 5mm, while the mean is around 10mm. These results indicate that the Hausdorff distance in WT predicted by TensorMixip model is generally about 5mm, but is relatively high for few abnormal data. Thus bring down the average prediction accuracy of the model in boundary of WT. The Hausdorff distance of TC is basically below 10mm, and that of ET is basically between 1mm~6mm, which indicates that the TensorMixup model can accurately identify the boundary of ET region.

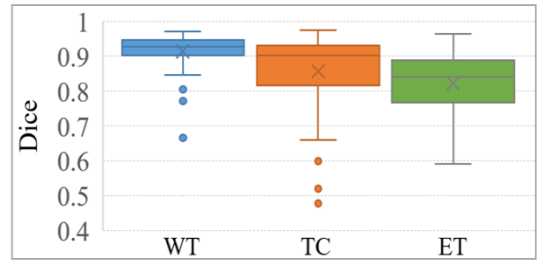

(a)

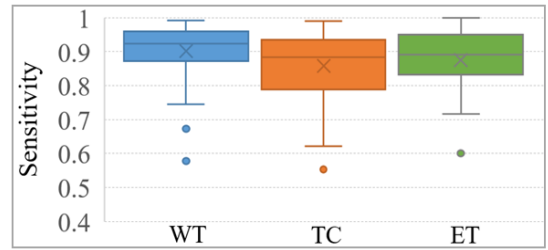

(b)

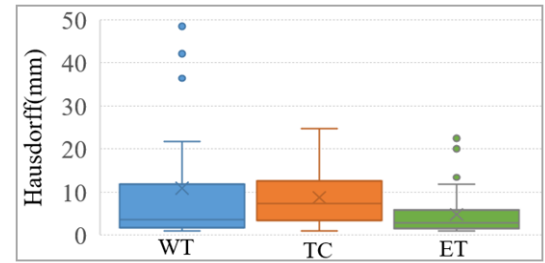

(c)

Fig. 6 Boxplot results of the proposed TensorMixup model on mean (a) Dice scores, (b) Sensitivities, and (c) Hausdorff distance on BraTs2019 testing dataset. The value of the line segment in each box represents the median of all the values in this box, and the value of the symbol '×' in each box represents the mean of all the values in this box.

In this study, TensorMixup, Mixup and basic augmentation method (Base) model were used to segment the brain MRI data, and the segmentation results on sample named BraTS_TCIA01_231_1 in BraTs2019 testing dataset were shown in Fig. 7. The three models were not exposed to the sample ever.

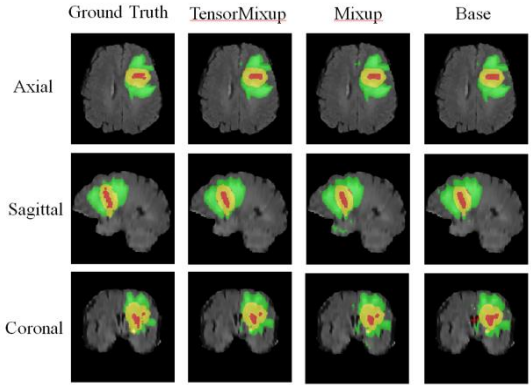

Fig. 7 Brain tumor segmentation results of BraTS_TCIA01_231_1 using three models. The Column 1 represents the raw Flair modal slices in Axial, sagittal and coronal plane. The Column 2-4 represents the segmentation results using TensorMixup, Mixup and Base model respectively. Green and yellow parts are the areas of edema and ET respectively, and red denotes the areas of non-enhancing tumor and necrosis.

There are small scattered connected regions outside the tumor regions in some segmentation results, which are results of mis-segmentation by the model and are called as false positive regions. Compared with Mixup and Base model, TensorMixup model can obtain more accurate segment results. In addition, the segmentation results of Base model in sagittal and coronal plane showed that the partly boundary and inside of enhancing tumor were wrongly segmented as necrotic tissues, and this model predicted two necrotic areas in the original image as a connected area. While TensorMixup model can identify the ET regions more accurately and its segmentation results have more smooth boundary, clear texture, and only tiny false positive areas. So it is more suitable for clinical application.

### 3.3.2 Effectiveness of Two Improvements in TensorMixup

In this study, two improvements including the use of tensor and the mixing of ROI were made based on traditional Mixup. For verifying the effectiveness of two improvements, the three mixing algorithms were used to train 3DUNet respectively on BraTs2015 dataset. The three methods are respectively that, using scalar to mix whole images (scalar+images), i.e. traditional Mixup, using scalar to mix ROIs (scalar+ROIs), using tensor to mix ROIs (tensor+ROIs), i.e. the proposed TensorMixup. Each model was trained with 190 epochs, and the evaluation results were shown in Tab. 3.

Tab. 3 Results of three models using different mixing algorithms in BraTs2015 testing dataset

| Augmentation method | Mean Dice (%) | | | Mean Sensitivity (%) | | | Mean Hausdorff (mm) | | |
|---|---|---|---|---|---|---|---|---|---|
| | WT | TC | ET | WT | TC | ET | WT | TC | ET |
| Scalar + images (Mixup) | 90.44 | 86.02 | 81.24 | 91.23 | 85.27 | 84.30 | 16.58 | 23.02 | 22.02 |
| scalar+ROIs | **91.04** | 86.72 | 82.68 | 90.37 | **88.16** | 82.97 | **12.21** | **11.15** | **8.54** |
| **tensor+ROIs (TensorMixup)** | 89.68 | **88.05** | **83.14** | **92.02** | 86.14 | **86.58** | 20.17 | 11.53 | 9.57 |

Compared with the results of scalar+ROIs and tensor+ROIs, in Tab. 3, it is found that the Dice scores of the tensor+ROIs method in TC and ET is higher than those of scalar+ROIs method by 1.33% and 0.46% respectively, which indicates that using a tensor to mix is beneficial to improving the segmentation accuracy of TC and ET. Compared with scalar+images method, The Dice scores of scalar+ROIs method in three regions increase by 0.6%, 0.7% and 1.44% successively. It demonstrates that mixing ROIs is helpful to some certain extent, and is more conducive to improving the segmentation performance of the model. In addition, In Tab. 3 the Hausdorff distances of scalar+ROIs method in three regions are lower than those of the other two compared methods, which shows that scalar+ROIs method can segment the boundary of the lesion region more accurately. This may also be one of the reasons why the scalar+ROIs method can predict WT region with a Dice score of 91.04%. In conclusion, the two main improvements in TensorMixup have a promoting effect on the segmentation performance of model.

### 3.3.3 Comparison with Other Methods

In order to further verify the performance of the proposed algorithm on the brain tumor segmentation task, in this paper TensorMixup is compared with state-of-the-art mixing algorithms such as CutMix [25] and Tumor Mixup [10] which are used to train 3DUNet model respectively on BraTs2019 dataset. The mean Dice scores of three models in WT, TC and ET were shown in Tab. 4.

Tab. 4 Comparison results with other algorithms in mean Dice index (%) on BraTs2019 testing dataset

| Method | WT | TC | ET |
| --- | --- | --- | --- |
| CutMix | 87.56 | 80.37 | 76.96 |
| Tumor Mixup | 87.73 | 80.13 | 78.35 |
| **TensorMixup** | **91.32** | **85.67** | **82.20** |

The results show that the segmentation accuracy of TensorMixup model is higher than those of the CutMix and Tumor Mixup models, which indicates that the proposed algorithm has greater advantages. The Dice scores of the two compared algorithms on TC and ET are obviously lower than that of TensorMixup method, because the location information of tumors in original images was not considered in the mixing process of the two compared algorithms. Thus destroying the topological structure of tumors in the synthetic image is destroyed, and the segmentation accuracy of the model is affected.

## 4 Conclusion

For improving the segmentation accuracy of brain tumors and making full use of multi-modal information of MRI data, a brain tumor segmentation algorithm, called TensorMixup, was proposed in this paper. Firstly, tumor image patches and their one-hot encoded labels are obtained from two MRI brain images with the same modalities. Then a tensor $\Lambda$ is used to mix the two image patches, and a matrix $\Lambda^*$ which is mapped from $\Lambda$ is used to mix the two one-hot encoded label sequences, so as to synthesize the new image and its one-hot encoded label, which could be used for training model and alleviating the problem of insufficient data in brain tumor segmentation task. The experimental results show that the Dice scores of the proposed algorithm in WT, TC and ET can reach 91.32%, 85.67% and 82.20% respectively, which are better than those of the model trained by basic data augmentation methods and traditional Mixup method respectively. The above results prove that the application of the proposed TensorMixup in brain tumor segmentation tasks has the feasibility and effectiveness.

## Reference

[1] Kleihues P, Burger P C, Scheithauer B W. The New WHO Classification of Brain Tumors[J]. Brain Pathology, 1993, 3(3): 255-268.

[2] Jiang Z K, Lyv X G, Zhang J X, et al. Review of Deep Learning Methods for MRI Brain Tumor Image Segmentation[J]. Journal of Image and Graphics, 2020, 25(02): 215-228.

[3] Havaei M, Davy A, Warde-Farley D, et al. Brain Tumor Segmentation with Deep Neural Networks[J]. Medical Image Analysis, 2017, 35: 18-31.

[4] Shen H, Zhang J and Zheng W. Efficient Symmetry-Driven Fully Convolutional Network for Multimodal Brain Tumor Segmentation[C]//2017 IEEE International Conference on Image Processing. IEEE, 2017: 3864-3868.

[5] Ronneberger O, Fischer P, Brox T. U-Net: Convolutional Networks for Biomedical Image Segmentation[C]//International Conference on Medical Image Computing and Computer-Assisted Intervention. Springer, Cham, 2015: 234-241.

[6] Dong H, Yang G, Liu F, Mo Y and Guo Y. Automatic Brain Tumor Detection and Segmentation Using U-Net Based Fully Convolutional Networks[C]//Annual Conference on Medical Image Understanding and Analysis. Springer, Cham, 2017: 506-517.

[7] Beers A, Chang K, Brown J, Sartor E, Mammen C P, Gerstner E, Rosen B and Kalpathy-Cramer J. Sequential 3D U-Nets for Biologically-Informed Brain Tumor Segmentation [J]. arXiv preprint arXiv: 1709.02967, 2017.

[8] Banerjee S, Mitra S, Shankar B U. Multi-planar Spatial-ConvNet for Segmentation and Survival Prediction in Brain Cancer[C]//International MICCAI Brain lesion Workshop. Granada, Spain, 2018: 94-104.

[9] Wang G T, Li W, Ourselin S, et al. Automatic Brain Tumor Segmentation Using Cascaded Anisotropic Convolutional Neural Networks[C]//Proceedings of the 3rd International Brainlesion: Glioma, Multiple Sclerosis, Stroke and Traumatic Brain Injuries. Quebec, Canada, 2017: 178-190.

[10] Li X Y, Luo G N, Wang K Q. Multi-step Cascaded Networks for Brain Tumor Segmentation[C]//International MICCAI Brainlesion Workshop. Springer, Cham, 2019: 163-173.

[11] Mckinley R, Meier R, Wiest R. Ensembles of Densely-Connected CNNs with Label-Uncertainty for Brain Tumor Segmentation[C]//International MICCAI Brainlesion Workshop. Granada, Spain, 2018: 456-465.

[12] Cirillo M D, Abramian D, Eklund A. What is the Best Data Augmentation Approach for Brain Tumor Segmentation using 3D U-Net?[J]. 2020.

[13] Zhang H Y, Cisse M, Dauphin Y N. Mixup: Beyond Empirical Risk Minimization[J]. arXiv preprint arXiv: 1710.09412, 2017.

[14] Zach E, Felix B, Sebastien O. Improving Data Augmentation for Medical Image Segmentation[C]//In International Conference on Medical Imaging with Deep Learning. 2020.

[15] Yin P, Hu Y, Liu J, et al. The Tumor Mix-Up in 3D U-Net for Glioma Segmentation[C]//International MICCAI Brainlesion Workshop. Springer, Cham, 2019: 266-273.

[16] Guo H. Nonlinear Mixup: Out-Of-Manifold Data Augmentation for Text Classification[C]//Proceedings of the AAAI Conference on Artificial Intelligence. AAAI, 2020, 34(4): 4044-4051.

[17] Menze B H, Jakab A, Bauer S, et al. The Multimodal Brain Tumor Image Segmentation Benchmark (BRATS)[J]. IEEE Transactions on Medical Imaging, 2014, 34(10): 1993-2024.

[18] Chapelle O, Weston J, Bottou L, et al. Vicinal Risk Minimization[J]. Advances in Neural Information Processing Systems, 2001: 416-422.

[19] Sun L C, Xia C Y, Yin W P, et al. Mixup-Transfomer: Dynamic Data Augmentation for NLP Tasks[C]//Proceedings of the 28th International Conference on Computational Linguistics. Barcelona, Spain, 2020: 3436-3440.

[20] Guo H, Mao Y, Zhang R. Mixup as locally linear out-of-manifold regularization[C]//Proceedings of the AAAI Conference on Artificial Intelligence. AAAI, 2019, 33(01): 3714-3722.

[21] Isensee F, Kickingereder P, Wick W, et al. Brain Tumor Segmentation and Radiomics

Survival Prediction: Contribution to the BraTS 2017 Challenge[C]//International MICCAI Brainlesion Workshop. Springer, Cham, 2017: 287-297.

[22] Xu B, Wang N, Chen T, et al. Empirical Evaluation of Rectified Activations in Convolutional Network[J]. arXiv preprint arXiv: 1505.00853, 2015.

[23] Kingma D, Ba J. Adam: a Method for Stochastic Optimization[C]//Proceedings of the 3rd International Conference on Learning Representations. San Diego, USA, 2015: 1-15.

[24] LI H. Statistical Learning Methods[M]. The Second Edition. Beijing: Tsinghua University Press, 2019: 23-24.

[25] Yun S, Han D, Oh S J, et al. Cutmix: Regularization Strategy to Train Strong Classifiers with Localizable Features[C]//Proceedings of the IEEE International Conference on Computer Vision. IEEE, 2019: 6023-6032.